\documentclass[conference]{IEEEtran}
\usepackage{cite}
\usepackage{amsmath,amssymb,amsfonts}
\usepackage{algorithmic}
\usepackage{graphicx}
\usepackage{textcomp}
\usepackage{xcolor}
\usepackage{multirow}
\usepackage{makecell}
\usepackage{wrapfig}
\usepackage{balance}
\usepackage[hidelinks]{hyperref}
\def\BibTeX{{\rm B\kern-.05em{\sc i\kern-.025em b}\kern-.08em
    T\kern-.1667em\lower.7ex\hbox{E}\kern-.125emX}}
\begin{document}

\title{An approach to improving\\sound-based vehicle speed estimation}

\author{
\IEEEauthorblockN{1\textsuperscript{st} Nikola Bulatovi\'c}
\IEEEauthorblockA{\textit{Faculty of Electrical Engineering} \\
\textit{University of Montenegro}\\
Podgorica, Montenegro \\
nbulatovic@ucg.ac.me}

\and

\IEEEauthorblockN{2\textsuperscript{nd} Slobodan Djukanovi\'c}
\IEEEauthorblockA{\textit{Faculty of Electrical Engineering} \\
\textit{University of Montenegro}\\
Podgorica, Montenegro \\
slobdj@ucg.ac.me}
}

\maketitle

\begin{abstract}

We consider improving the performance of a recently proposed sound-based
vehicle speed estimation method. In the original method, an intermediate feature,
referred to as the modified attenuation (MA), has been proposed for both
vehicle detection and speed estimation. The MA feature maximizes at the instant
of the vehicle's closest point of approach, which represents a training label
extracted from video recording of the vehicle's pass by.
In this paper, we show that the original labeling approach is suboptimal
and propose a method for label correction.
The method is tested on the VS10 dataset, which contains $ 304 $ audio-video recordings
of ten different vehicles. The results show that the proposed label correction
method reduces average speed estimation error from $ 7.39 $ km/h to $ 6.92 $ km/h.
If the speed is discretized into $ 10 $ km/h classes,
the accuracy of correct class prediction is improved from $ 53.2\% $ to $ 53.8\% $,
whereas when tolerance of one class offset is allowed, accuracy
is improved from $ 93.4\% $ to $ 94.3\% $.

\end{abstract}


\section{Introduction}
\label{Intro}

Traffic analysis and planning represent indispensable parts of
an intelligent transportation system. Large volumes of traffic data
enable significant improvements in the performance of transportation,
traffic safety and automatic traffic monitoring (TM) \cite{won2020survey}.
The TM data are used for valuable information extraction \cite{shokravi2020review},
which may include vehicle count \cite{djukanovic2020robustcount,djukanovic2021neuralcount},
shape \cite{velazquez2018vehicledetection,simoncini2018vehicle},
speed \cite{djukanovic2021vehiclespeed,bulatovic2022melspectrogram}, 
acceleration \cite{sun2013vehicle}, 
type \cite{manzoor2018vehicle,ghassemi2019vehicle},
plate number \cite{biglari2017cascaded} and may be used to
predict road accidents \cite{almaadeed2018automatic,foggia2015audio}.

Diverse sensing technologies are utilized in modern TM systems.
Some of them involve induction loops, piezoelectric and magnetic sensors,
ground and aerial cameras, acoustic sensors, Doppler radar and LIDAR
\cite{won2020survey}. A study conducted in \cite{naphade2019challenge}
describes recent advances in computer vision applied in multi-camera
vehicle tracking and identification, as well as traffic anomaly detection.
TM based solely on vision data is often not sufficient due to sensitivity
of image sensors to adverse weather conditions, lighting changes,
vehicle headlights and shadows \cite{won2020survey,morris2008survey,
foggia2015audio,crocco2016audio}.

Advantages of the acoustic-based TM over the other technologies include
low price, low energy consumption and low storage space required.
Acoustic sensors are easier to install and maintain and are not affected
by visual occlusions and lighting variations. They are also less disturbing
to drivers' behavior and have fewer privacy issues \cite{won2020survey},
\cite{wilson2010speedcameras}. Current acoustic TM methods, described in
\cite{djukanovic2021vehiclespeed}, can be divided into single-microphone
\cite{quinn1995doppler,couvreur1997doppler,cevher2008vehiclespeed,
barnwal2013doppler,kubera2019speedchanges,koops2015speedgear,
giraldo2016vehiclespeed,goksu2018vehiclespeed} and microphone-array approaches
\cite{lopez2004vehiclespeed, lo2000targetmotion, marmaroli2013twoaxle}.

One of important concerns in machine learning and deep learning applications 
for acoustic vehicle detection and speed estimation is the lack of high-quantity
and high-quality annotated data \cite{cevher2008vehiclespeed,barnwal2013doppler,
giraldo2016vehiclespeed,koops2015speedgear,marmaroli2013twoaxle,goksu2018vehiclespeed}.
Furthermore, deep neural networks tend to overfit to corrupted labels, which results
in poor generalization on the test dataset \cite{zhang2021understanding}. 
This is why robust training, in the presence of noisy labels,
is an important task in supervised learning applications \cite{song2020learning}.

In this paper, we consider improving the accuracy of single-sensor acoustic vehicle 
detection and speed estimation, proposed in \cite{djukanovic2021vehiclespeed}.
The original method introduced an analytical speed-dependent feature, referred to as
the modified attenuation (MA), which maximizes at the instant of the vehicle's
closest point of approach (CPA). The CPA instant is represented by a training label
manually extracted from video recording of the vehicle's pass by.
This labeling procedure is not optimal due to inconsistent camera view angles at installation sites.
Time offset, caused by suboptimal labeling approach, is referred to as the labeling noise.
This paper addresses label correction aimed to improve the accuracy 
of vehicle speed estimation. To that end, we propose to shift the ground-truth CPA instants
by amount obtained from the test phase of the MA prediction.
This approach yields notable speed estimation accuracy improvement.

We first overview the original speed estimation approach in Section \ref{SpeedEstimation}
and then propose label correction method in Section \ref{NoisyLabelsCorrection}.
Experimental results are given in Section \ref{Experiment} and conclusions
are drawn in Section \ref{Conclusion}.

\section{Speed Estimation}
\label{SpeedEstimation}

\subsection{Data}
\label{Dataset}

Audio recordings dataset from \cite{djukanovic2021vehiclespeed}, 
referred to as VS10, is used to train and test the speed estimation method. 
Recordings were captured with a GoPro Hero5 Session camera and divided into
$ 304 $ single-vehicle and $ 71 $ environmental-noise audio-video files.
Ten different vehicles are included, as reported in Table I in 
\cite{djukanovic2021vehiclespeed}. Vehicles' speeds range
from $ 30 $ km/h to $ 105 $ km/h. Audio data are extracted from video recordings.
Audio files are clipped to $ 10 $ seconds with $ 44100 $ Hz sampling rate
and saved in WAV format with $ 32 $-bit float PCM.
Dataset labels contain the vehicle's speed and its pass-by time (CPA instant).
The pass-by instant is obtained manually by identifying
the time frame of a vehicle starting to exit the camera view.

\subsection{Acoustic features}
\label{SpeedFeature}

In this paper, we use modified attenuation, originally proposed in \cite{djukanovic2021vehiclespeed}
for both vehicle detection and speed estimation. The MA feature represents a modification of the amplitude
attenuation factor of the sound signal \cite{couvreur1997doppler}.

The MA feature is defined as follows:
\begin{equation}
	\label{ModifAtten}
	\eta(t)=\frac{\alpha}{\beta v^2(t_{\text{CPA}}-t)^2 + d_{\text{CPA}}^2}
\end{equation}
where $ v $ denotes the speed of vehicle, $ t $ is time,
$ t_\text{CPA} $ is the CPA time instant, and $ d_{\text{CPA}} $
denotes the vehicle distance at CPA. Vertical and horizontal extents
of $ \eta(t) $ are adjusted by $ \alpha $ and $ \beta $, respectively.
Appropriate settings of $ \alpha $ and $ \beta $ are outlined in
Section \ref{Implementation}.

MA is predicted in a supervised fashion from the log-mel spectrogram (LMS) of input audio.
LMS represents a very popular feature in acoustic classification applications
\cite{serizel2018acousticfeatures} and it proved very reliable in vehicle detection
and speed estimation \cite{djukanovic2020robustcount,djukanovic2021neuralcount,
bulatovic2022melspectrogram}.

\subsection{Method}
\label{SpeedEstimationMethod}

\begin{figure}[t]
    \centerline{\includegraphics[scale=0.18]{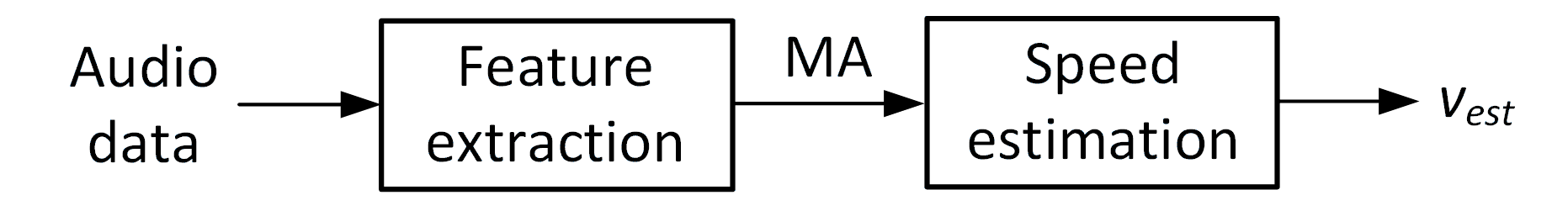}}
    \caption{Block diagram of the speed estimation method.}
    \label{figblockdiagram}
\end{figure}

\begin{figure}[htbp]
    \centerline{\includegraphics[scale=0.68]{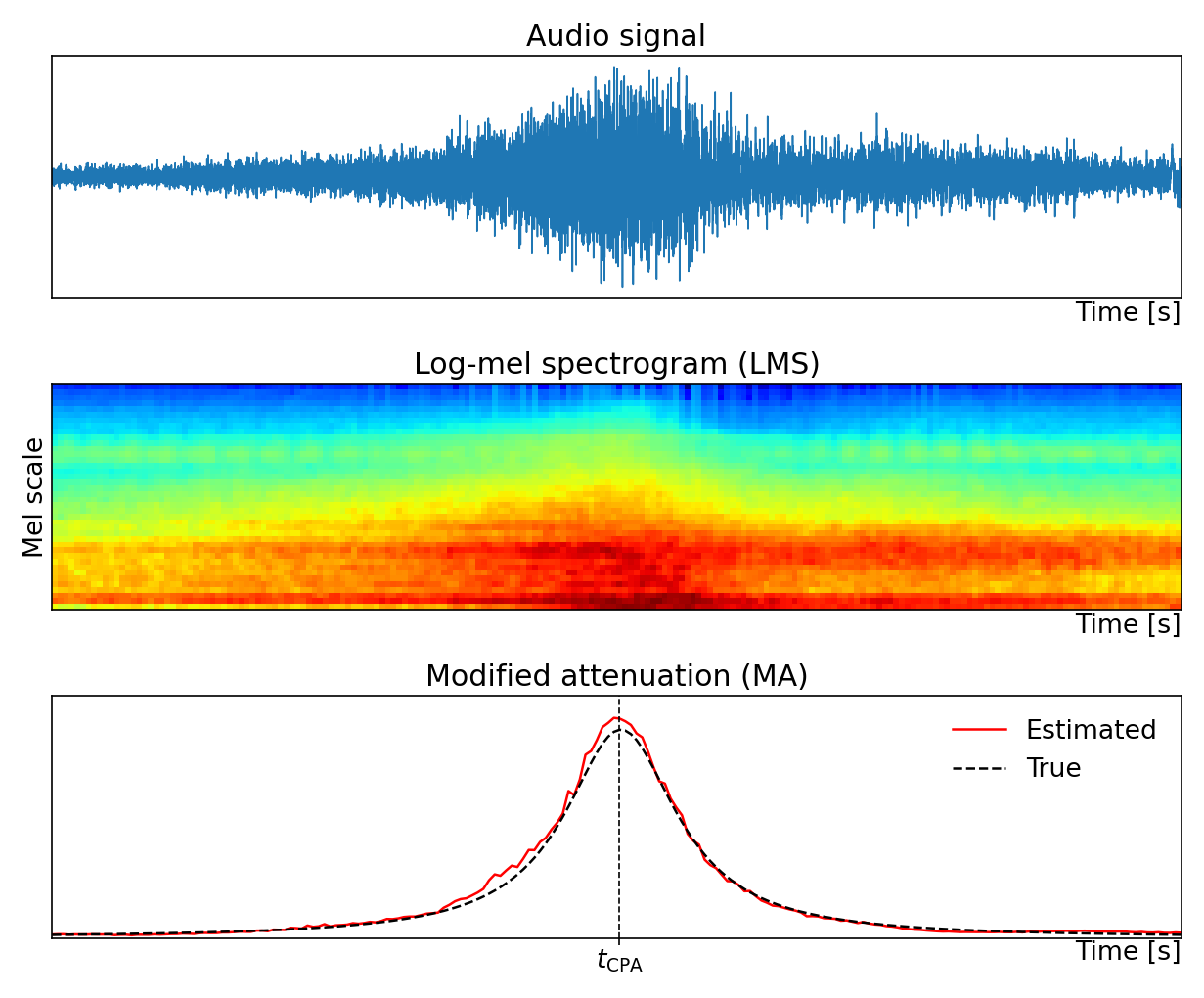}}
    \caption{{\it{Top:}} Original audio signal of vehicle's pass-by (speed $ 80 $ km/h).
	{\it{Middle:}} Log-mel spectrogram of audio signal. 
	{\it{Bottom:}} True and estimated MA features.}
    \label{figfeatures}
\end{figure}

The vehicle speed estimation method is introduced in \cite{djukanovic2021vehiclespeed}
and outlined in Fig. \ref{figblockdiagram}. In the first step, input audio signal is transformed
to its short-time power spectrum representation, namely LMS. LMS remaps original
audio frequencies to their logarithmic mel scale. Next, the MA feature regression
is performed using a fully-connected deep neural network (DNN).
One example of MA prediction is presented in Fig. \ref{figfeatures}.
The second step comprises speed estimation from the predicted MA.
The vehicle's pass-by time, denoted as $ t_{\text{CPA}} $ and estimated by maximizing
the MA profile, is used to detect the vehicle and select an appropriate interval
of predicted MA samples for speed estimation. Finally, the speed is estimated
using the selected MA samples. As in \cite{djukanovic2021vehiclespeed},
$ \varepsilon $-support vector regression ($ \varepsilon $-SVR) \cite{chang2011libsvm} is used
in speed estimation due to relatively small size of the dataset.

\subsection{Implementation details}
\label{Implementation}

\subsubsection{Features for MA regression}

Original audio signals are clipped to $ 10 $ seconds with the sampling rate
of $ 44100 $ Hz. LMS of audio is calculated using the short-time Fourier transform (STFT).
STFT is calculated with $ 4096 $-sample Hamming window and $ 1105 $-sample hop rate, 
which results in $ 400 $ time frames of STFT \cite{stankovic2014timefrequency}. For each time frame, 
$ N_{mel}=40 $ mel coefficients are computed, within the frequency interval of $ [0,16 \text{ kHz}] $.

The MA feature (\ref{ModifAtten}) is calculated with $ \alpha=v $, $ \beta=0.05 $
and $ d_{\text{CPA}}=1.5 $ m. The MA value, at instant $ t $, is estimated using
a window of LMS features comprising $ Q=25 $ time frames (stride of $ 3 $),
centered at $ t $. This setting yields $ M=Q \times N_{mel}=1000 $ LMS input features.
Fully-connected DNN consists of $ 1000 $--$ 200 $--$ 50 $--$ 10 $--$ 1 $ neurons per layer.
DNN is implemented using mean squared error loss, ReLU activation
(linear activation in the last layer), $ L2 $ kernel regularization 
with factor $ 10^{-3} $ and $ 200 $ training epochs.

\subsubsection{Speed estimation}
\label{SpeedEstimationImplementation}

Optimal $ \varepsilon $-SVR parameters, namely $ C=10 $ and $ \varepsilon=0.1 $,
are obtained by a grid search. A vector of $ 73 $ adjacent MA coefficients, 
centered at $ t_{\text{CPA}} $, represents input features of the $ \varepsilon $-SVR block.

The speed estimation model is evaluated with $ 10 $-fold cross-validation.
Nine folds (vehicles) are used in training and validation, while the remaining
fold is put aside for testing. Training and validation datasets are split
in a $80\%$-$20\%$ fashion, as described in Section II-B in \cite{djukanovic2021vehiclespeed}.
The training procedure is repeated $ 20 $ times and the results are averaged.

\section{Noisy Labels Correction}
\label{NoisyLabelsCorrection}

Precise prediction of $ t_{\text{CPA}} $ is of pivotal importance for accurate
speed estimation. As mentioned in Section \ref{SpeedEstimationMethod}, $ t_{\text{CPA}} $
is estimated by maximizing the predicted MA feature. The vehicle's speed is then
estimated from a window of MA samples centered at $ t_{\text{CPA}} $
(see Section \ref{SpeedEstimationImplementation}).

\begin{figure}[t]
    \centerline{\includegraphics[scale=0.72]{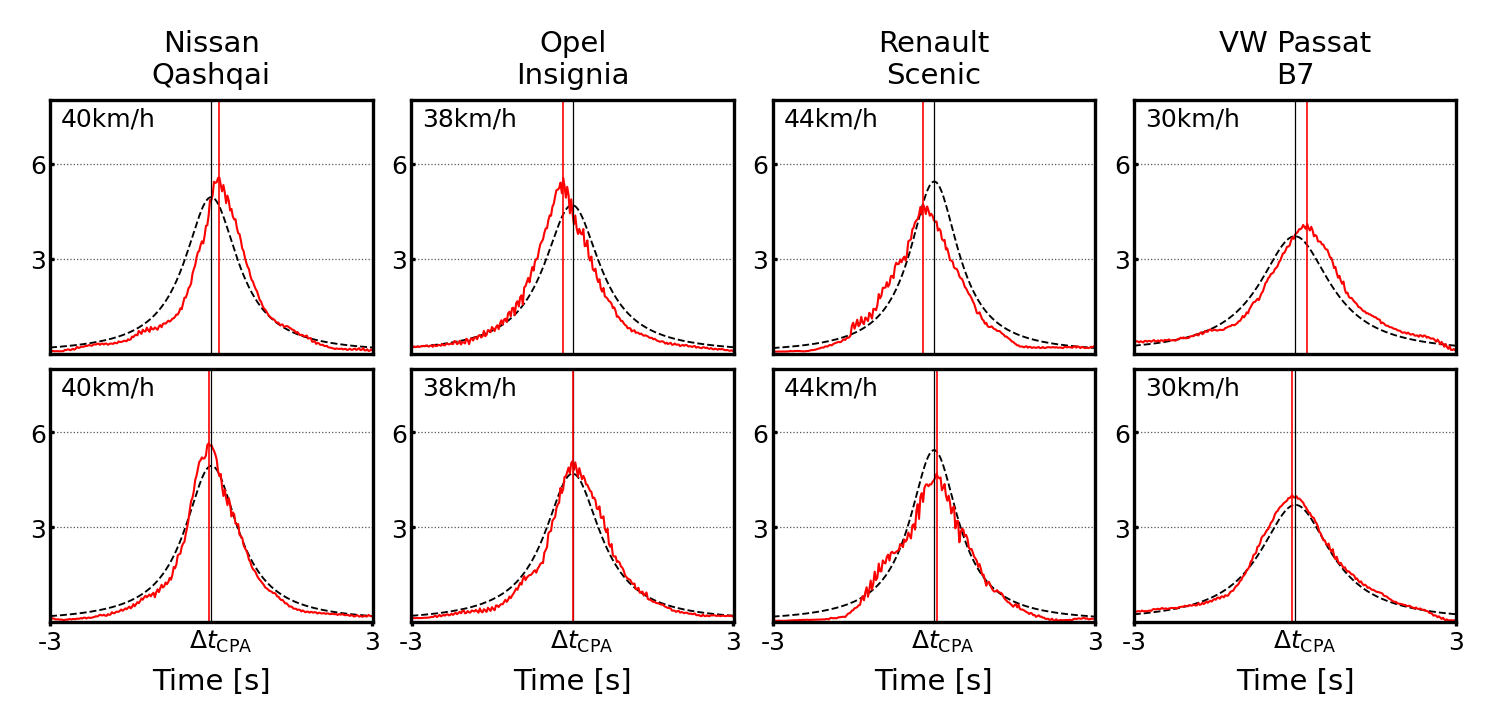}}
    \caption{Predicted MA features (solid red) compared to their ground-truth (dashed black).
	The offset between the ground-truth and the predicted CPA instants (vertical lines),
	$ \Delta t_{\text{CPA}} $, represents labeling noise.
	{\it{Top:}} MA predictions using original labels.
	{\it{Bottom:}} MA predictions after label correction is implemented.}
    \label{figMApredicted}
\end{figure}

The VS10 recordings, described in Section \ref{Dataset}, were captured with a camera installed
at the side of a local road close to Podgorica, Montenegro (see Fig. 1 in 
\cite{djukanovic2021vehiclespeed}).
Both incoming directions are present in the dataset.
Camera positions and view angles differ slightly for each recording.
Various angles of the camera view, with respect to the vehicle's trajectory,
lead to imprecise $ t_{\text{CPA}} $ labels in the dataset.
The predicted MA profiles (testing phase) are shifted with respect to the ground-truth ones
by $ \Delta t_{\text{CPA}} $, as depicted in Fig. \ref{figMApredicted} (top).
This notable shift between the ground-truth and the predicted MA values motivated us
to correct the $ t_{\text{CPA}} $ labels by taking into account the resulting MA predictions
on the test dataset.

Generalization of the proposed MA regression is degraded due to imprecise $ t_{\text{CPA}} $ labeling.
Since the true $ t_{\text{CPA}} $ value cannot be determined precisely by analyzing the video-recording
(camera view angle is not known), we propose to correct the ground-truth $ t_{\text{CPA}} $
by using the predicted MA profiles of the test data.
More precisely, for each audio file, the corrected $ t_{\text{CPA}} $ is calculated as a median $ t_{\text{CPA}} $
over $ 20 $ runs of the proposed MA regression model. For the $ t_{\text{CPA}} $ correction,
we consider only the test results of MA regression. Since the test data correspond to a vehicle
not included in the training phase, the regression model does not overfit to the test data.
After the labels are corrected, the model is trained again. Improvements in $ t_{\text{CPA}} $ predictions,
due to the proposed label correction, are depicted in Fig. \ref{figMApredicted} (bottom).
Speed estimation improvements are presented in Section \ref{Experiment}.

\section{Experiments and Results}
\label{Experiment}

\begin{figure}[t]
    \centerline{\includegraphics[scale=0.8]{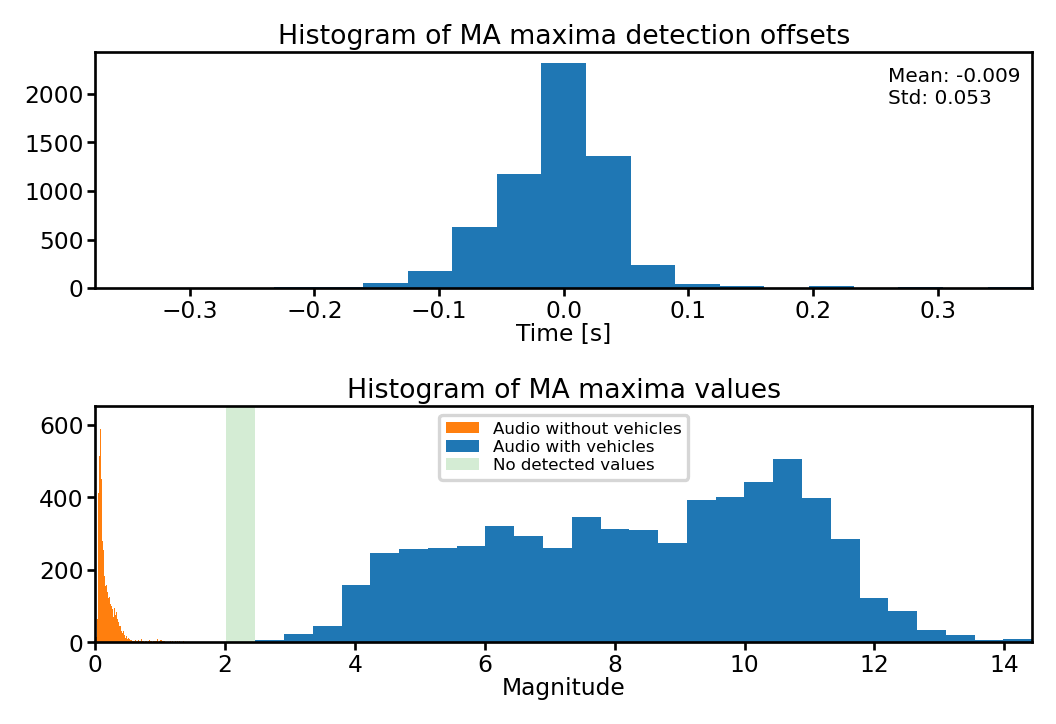}}
    \caption{{\it{Top:}} Histogram of MA maxima detection offsets on the test dataset.
	{\it{Bottom:}} Histogram of MA maxima for audio with vehicles (blue) and without vehicles (orange).
	Green rectangle separates the two cases.}
    \label{figdetection}
\end{figure}

The proposed method is evaluated in terms of vehicle detection and vehicle speed estimation accuracy. 
Since vehicle detection is defined as estimation of the $ t_{\text{CPA}} $ instant,
we calculate detection error as a difference between the true and the estimated MA maxima positions.
Distribution (histogram) of MA maxima detection offsets of the test data
are presented in Fig. \ref{figdetection} (top). The histogram comprises the test results
from $ 20 $ DNN training iterations. If the error is modeled as a normal random variable,
its mean value and standard deviation are improved with respect to \cite{djukanovic2021vehiclespeed},
from $ -0.016 $ to $ -0.009 $ and from $ 0.065 $ to $ 0.053 $, respectively.
In Fig. \ref{figdetection} (bottom), we compared histograms of the MA maxima values in two 
scenarios, with vehicle (blue histogram) and without vehicle (orange histogram) passing by the microphone.
The width of the green rectangle area, which separates the vehicle and no-vehicle cases, increased
from $ 0.30 $, obtained in \cite{djukanovic2021vehiclespeed}, to $ 0.45 $. This area represents
the optimal range to set the MA magnitude threshold for accurate vehicle detection.

\begin{table}[t]
	\centering
	\caption{Speed estimation RMSE}
	\begin{tabular}{ l c }
		\hline\hline
		\multicolumn{1}{c}{Vehicle} & \multicolumn{1}{c}{RMSE [km/h]} \\ \hline
		Citroen C4 Picasso & $6.40$ \\ \hline
		Mazda 3 Skyactive & $8.57$ \\ \hline
		Mercedes AMG 550 & $6.89$ \\ \hline
		Nissan Qashqai & $6.21$ \\ \hline
		Opel Insignia & $4.25$ \\ \hline
		Peugeot 307 & $6.65$ \\ \hline
		Peugeot 3008 & $7.37$ \\ \hline
		Renault Captur & $5.81$ \\ \hline
		Renault Scenic & $10.52$ \\ \hline
		VW Passat B7 & $6.50$ \\ \Xhline{1pt}
		Average & $6.92$ \\ \hline\hline
	\end{tabular}
	\label{Tab1}
\end{table}

\begin{table}[t]
	\centering
	\caption{Speed classification accuracy with $ \Delta $ denoting the distance from the true class}
	\begin{tabular}{l r r r r r}
		\hline\hline
		\multicolumn{1}{c}{Vehicle} & \multicolumn{1}{c}{$\Delta=0$} &
		\multicolumn{1}{c}{$|\Delta|=1$} & \multicolumn{1}{c}{$|\Delta|=2$} &
		\multicolumn{1}{c}{$|\Delta|>2$} & \multicolumn{1}{c}{$|\Delta|\leq 1$} \\ \hline
		Citroen C4 Picasso & $62.6\%$ & $28.9\%$ & $8.5\%$ & $0.0\%$ & $91.5\%$ \\ \hline
		Mazda 3 Skyactive & $43.9\%$ & $48.3\%$ & $6.2\%$ & $1.6\%$ & $92.2\%$ \\ \hline
		Mercedes AMG 550 & $50.2\%$ & $48.5\%$ & $1.3\%$ & $0.0\%$ & $98.7\%$ \\ \hline
		Nissan Qashqai & $42.4\%$ & $55.2\%$ & $2.4\%$ & $0.0\%$ & $97.6\%$ \\ \hline
		Opel Insignia & $69.8\%$ & $30.2\%$ & $0.0\%$ & $0.0\%$ & $100.0\%$ \\ \hline
		Peugeot 307 & $62.2\%$ & $31.0\%$ & $6.6\%$ & $0.2\%$ & $93.3\%$ \\ \hline
		Peugeot 3008 & $48.7\%$ & $41.8\%$ & $8.5\%$ & $1.0\%$ & $90.5\%$ \\ \hline
		Renault Captur & $61.7\%$ & $33.3\%$ & $5.0\%$ & $0.0\%$ & $95.0\%$ \\ \hline
		Renault Scenic & $44.9\%$ & $42.3\%$ & $8.7\%$ & $4.1\%$ & $87.1\%$ \\ \hline
		VW Passat B7 & $52.0\%$ & $45.4\%$ & $2.6\%$ & $0.0\%$ & $97.4\%$ \\ \Xhline{1pt}
		Average & $53.8\%$ & $40.5\%$ & $5.0\%$ & $0.7\%$ & $94.3\%$ \\ \hline\hline
	\end{tabular}
	\label{Tab2}
\end{table}

Speed estimation accuracy is evaluated using root-mean-square error (RMSE)
\begin{equation}\label{RMSE}
	\text{RMSE}=\sqrt{\frac{1}{L}\sum\nolimits_{l=1}^{L}(v_{l}^{est}-v_{l}^{true})^2},
\end{equation}
with $ v_{l}^{est}$ and $v_{l}^{true} $ representing the estimated and true speeds
of the $ l $-th measurement, respectively, whereas $ L $ is the total number of speed measurements.
Average RMSE values, presented in Table \ref{Tab1}, demonstrate the effect of $ t_{\text{CPA}} $
label correction in the training dataset. Compared with \cite{djukanovic2021vehiclespeed},
the total average RMSE value decreases from $ 7.39 $ km/h to $ 6.92 $ km/h. 
Vehicle-by-vehicle RMSE comparison with Table II in \cite{djukanovic2021vehiclespeed} shows
performance improvement for each vehicle except for Opel Insignia, where RMSE is slightly increased.

Speed estimation accuracy is also evaluated in a classification framework.
Namely, vehicle speeds are divided into $ 8 $ classes, starting from $ 25 $ km/h, with a step of $ 10 $ km/h.
Speed classification results are presented in Table \ref{Tab2}, with $ \Delta $ denoting the distance from the
true class. The presented results show improvements of average speed classification accuracies
in the cases of $ \Delta=0 $ (true and predicted classes coincide) and
$ |\Delta|\leq 1 $ (true and predicted classes are adjacent ones)
from $ 53.2\% $ to $ 53.8\% $ and from $ 93.4\% $ to $ 94.3\% $, respectively.
Note that the presented speed classification results of Renault Scenic and Mazda 3 Skyactive are 
highly affected by a background environmental noise during dataset acquisition.
On the other hand, Opel Insignia outperforms the other vehicles
with the outstanding $ |\Delta|\leq 1 $ accuracy of $ 100.0\% $.

\section{Conclusion}
\label{Conclusion}

In the paper, we considered how the test results of intermediate MA feature regression
could be used to improve coarse labels, obtained by visual scene inspection.
The proposed label correction method improved the average speed estimation error by $ 0.47 $ km/h.
When formulated as a classification problem, with the discretization interval of $ 10 $ km/h, 
the achieved accuracy improved $ 0.6\% $ for correct class prediction and $ 0.9\% $
in the case when one class offset is allowed.

Our future vehicle speed estimation research will focus on data-oriented approaches
such as data augmentation and extending the dataset with additional vehicles.

\bibliographystyle{IEEEtran}
\bibliography{paper}

\end{document}